# Tsunami and megathrust earthquake disaster prevention warnings: Real-time monitoring of the genesis processes with Physical Wavelets

Fumihide Takeda [1,2]

[1] Takeda Engineering Consultant Co, Hiroshima, Japan

[2] Earthquake Prediction Institute, Imabari, Japan

A megathrust earthquake genesis of 15 months with its tsunami genesis of the last 3 months provides a real-time disaster prevention warning and hazard mitigation measures leading up to the events.

**Abstract**

Japan's seismograph [1] and GPS [2] network provide comprehensive records on earthquakes (EQs) in the form of displacement time series data, but these records are noisy and stochastic. Physical Wavelets (PWs) are powerful mathematical operators that can define position (displacement), velocity, and acceleration on the stochastic time series data of EQ records [3, 4]. These operators have been used to define equations for megathrust and significant EQ genesis processes, which enable the prediction of EQ focuses, fault movements, sizes, and rupture times up to three months in advance of the EQ occurrence with accuracies within a day [3-6].

PWs have also been applied to illustrate the 2011 Tohoku megathrust EQ genesis process, including the tsunami genesis process of the last three months [4, 6]. Such illustrations could have provided real-time disaster prevention warnings and hazard mitigation measures leading up to the event. However, effectively informing societal decisions remains a significant challenge [4, 7, 8] that needs to be addressed.

**GPS stations in the Tohoku subduction zone**

Figure 1 shows the GPS stations in the Tohoku subduction zone. They provided a distribution of the Tohoku EQ's coseismic shifts, as shown in Fig. part a. Shift down, no change, and up are along the east coast, the ridge, and the west coast, respectively. The three GPS stations in Fig. part b, labeled as 1, 2, and 3, are referred to as the Onagawa station (the east coast), the Murakami station (the ridge, Top), and the Ryoutsu2 station (the west coast), respectively.

**GPS displacement time series data**

GPS stations provide each daily position whose relative change is non-differentiable displacement $d(c, j)$ at time $j$ (in days) where the geographic axis $c$ represents $E$ (eastward), $N$ (northward), and $h$ (upward) in right-handed coordinates $(E, N, h)$. The daily record of $d(c, j)$ is denoted by $\{c\}=\{d(c, 0), ., d(c, j), .\}$ that is non-differentiable due to environmental noises.

**Physical Wavelets**

Physical Wavelets (PWs) are powerful operators to define position (displacement), velocity, and acceleration on the stochastic time series data $\{c\}$ [3-5]. The operators are the displacement-defining operator $DDW(t − \tau)$, velocity-defining operator $VDW(t − \tau)$, and acceleration-defining operator $ADW(t − \tau)$ at time $\tau$. They satisfy the time reversal property for positions and their derivatives, enabling the definition of respective fundamental physical quantities. By taking the cross-correlation (or inner product in the case of vector representations) of each operator with the non-time-differentiable series $\{c\}$, we can define displacement $D(c, \tau)$, velocity $V(c, \tau)$, and acceleration $A(c, \tau)$ at time $\tau$ as follows:

$$D(c,\tau) = \int_{-\infty}^{+\infty} \{c\} \, DDW(t-\tau) \, dt = [1/(2w+1)] \sum_{j=-w}^{w} d(c, \tau + j), \qquad (1)$$

$$V(c,\tau) = \int_{-\infty}^{+\infty} \{c\} \, VDW(t-\tau) \, dt = [D(c, \tau + s/2) - D(c, \tau - s/2)]/s \qquad (2)$$

and

$$\begin{aligned} A(c,\tau) &= \int_{-\infty}^{+\infty} \{c\} \, ADW(t-\tau) \, dt = [V(c, \tau + s/2) - V(c, \tau - s/2)]/s \\ &= [D(c, \tau + s) - 2D(c, \tau) + D(c, \tau - s)]/s^2. \end{aligned} \qquad (3)$$

In Eq. (1), $D(c, \tau)$ is the average of $d(c, j)$ over a time interval of $j = \tau \pm w$. The time-reversal operation changes $\tau$ to $-\tau$ and confirms that $D(c, -\tau) = D(c, \tau)$, $V(c, -\tau) = -V(c, \tau)$, and $A(c, -\tau) = A(c, \tau)$. Thus, the operators exhibit time-differential properties for the differentiable $D(c, \tau)$ and $V(c, \tau)$ obtained from the non-differentiable $d(c, j)$ of $\{c\}$, which may have periodically fluctuating components and trends. The extraction of specific periodicities and trends is significant only if the mutual correlation between PWs and $\{c\}$ is strong. Eqs. (1), (2), and (3) represent low pass filtered, bandpass filtered, and another bandpass filtered physical quantity, respectively. The parameters $w$ and $s$ can be any integer used to filter out the selected frequency components and trends in $\{c\}$ to define $D(c, \tau)$, $V(c, \tau)$, and $A(c, \tau)$. The relationships between $D(c, \tau)$, $V(c, \tau)$, and $A(c, \tau)$ represent the equations of motion for the observed time series $\{c\}$, which is non-differentiable in time.

**Power monitoring**

We define a time-rate change of the kinetic energy (velocity squared) as the product of $V(c, \tau)$ and $A(c, \tau)$, which is the power, $PW(c, \tau)$. We reassign them as $V(c, j)$ and $A(c, j)$ at time $j (= \tau + w + s)$, defined for $A(c, \tau)$. The power is then calculated as $PW(c, j) = V(c, j) \times A(c, j)$. Monitoring the oceanic plate and the Tohoku crust's motions with $PW(c, j) \geq$ TH (a predetermined threshold) can detect any unexpected motion and its onset. The threshold level TH is set for detecting the unexpected motion, based on the observed $PW(c, j)$'s maximum amplitude during expected standard motions.

**The Tohoku megathrust EQ and tsunami genesis processes**

The Tohoku M9 EQ and the resulting tsunami were not caused by the elastic rebound of the east coast subsiding with the subducting oceanic plate coupled with the fault [3, 4]. Instead, they resulted from complex interactions between the subducting and overriding plates, which caused bulge-bending deformation over the Tohoku subduction zone. The bulge-bending deformation refers to the gradual bending of the overriding plate of Tohoku by the eastward continental plate-driving force, causing the over-riding crust of Tohoku to bulge upwards and downwards by a few millimeters. This deformation evolved from the expected regular deformation that had been occurring for the past three hundred years [4].

**Distinctive phases of the Tohoku crustal deformation**

The GPS observation detected the onset of the bulge-bending deformation, which grew in three distinct phases: an initial phase of 6 months with gradual subsidence of 1 to 2.8 mm across the Tohoku region, followed by a transitional phase of 6 months with further gradual subsidence of 3.3 mm on the east coast, and a final phase of 3 months with an upheaval growth of 1 to 3 mm across Tohoku until the 2011 EQ event.

**Regular phase (Before Jan 2010)**

Figure 2a illustrates the regular deformation of the dotted line over Tohoku with the subducting oceanic plate moving westward, causing the east coast to subside (represented by a westward arrow) and the overriding continental plate moving eastward, causing the west coast to move upward (represented by an eastward arrow). For instance, the Onagawa station (in Fig. 1) had a subsidence rate of 6 mm per year, while the Ryoutsu2 station (in Fig. 1) had an upward displacement rate of 1.5 mm per year.

**Initial phase (January 2010 to June 2010)**

One month prior to January 2010, the deformation property of the west coast changed from non-elastic to elastic, leading to the generation of a restoring force on the compressed west coast. This property change of the west coast marks the onset of an underlying bulge-bending deformation that persisted for 15 months, resulting in gradual subsidence and upheaval growth of a few millimeters over a 500 km stretch along the east and west coasts.

In January 2010, the westward-restoring force of the west coast, compressed by the overriding continental plate-driving force, caused a gradual subsidence of 1 to 2.8 mm across the Tohoku region. By June 2010, the subsidence had reached 1 mm on the west coast, 1.5 mm on the top, and 2.8 mm on the east coast. The east coast subsidence was a pre-transition process from regular to bulge-bending deformation.

**Transitional phase (June 2010 to November 2010)**

In June 2010, the transitional phase began with a further gradual subsidence of 3.3 mm on the east coast until the final phase, as illustrated in Fig. 2b. The subsidence firmly grasped the fault and started to pull down the subducting Pacific Plate on July 11, 2010, represented by the dotted arrow over the fault. The pulling action, caused by compressing the west coast, accelerated the westward movement, reaching the highest speed of 0.69 mm/day on December 22, 2010.

About a month before reaching the highest speed, the east coast bulge shifted to the final phase of upheaval growth. The linear upheaval growth of 1.2 mm over 115 days began on November 14, 2010. The Top began the upheaval growth on August 26, 2010, reaching 2.5 mm before the 2011 M9 EQ event. The gradual upheaval growth of 2.3 mm on the west coast began on October 29, 2010.

**Final phase (November 2010 to March 2011)**

In November 2010, the final phase began with the gradual upheaval growth of 1.2 mm on the east coast. The bulge-bending force with this growth decelerated the subducting Pacific Plate's westward movement, which halted by February 21, 2011, and remained motionless for another four days. On February 25, the bulge force reversed the westward plate motion, with the eastward speed reaching 0.06 mm/day just three days before the M9 EQ on March 11, 2011.

The final upheaval growth of 1.2 mm rapidly released a massive restoring force in the west coast, compressed elastically by the overriding plate-driving eastward force. This recoil restored the bulge-bent west and east coasts, ultimately decoupling the overriding and subducting plates and leading to the megathrust EQ and tsunami on March 11, 2011, as depicted by the dotted arrow in Fig. 2c. Throughout the transitional and final phases, Tohoku experienced an elastic compression of 10.0 mm on the west coast and a pulling movement of 13.2 mm on the east coast in the same westward direction as the subducting oceanic plate-driving westward force, persisting until the 2011 Tohoku EQ event on March 11, 2011.

**Tsunami genesis process**

As discussed in the reference [4], the relationship between averaged accelerations (forces) and displacements on the Tohoku subduction zone indicates that the compression of the west coast by the overriding plate-driving eastward force is elastic. The plate-driving eastward force simultaneously compressed (bulge-bent) the east coast westward across Tohoku, not the subducting plate-driving westward force. Thus, the Tohoku M9 earthquake and tsunami were not caused by the commonly suggested elastic rebound of the east coast compressed by the subducting plate-driving force coupled with the over-riding plate through the fault.

The overriding plate-driving eastward force compressed the west coast by +18.4 mm (eastward until the Tohoku M9 EQ on March 11, 2011), initiating the underlying bulge-bending deformation during the initial phase and generating the east coast's pulling action during the transitional and final phases with -13.2 mm (westward) movement. During the final phase, the upheaval growth of 1.2 mm resulted in the accumulation of lifting force along the east coast, which weakened the static frictional strength of the subduction interface. This weakening eventually caused the shear stress to exceed the frictional strength, decoupling the overriding and subducting plates. The decoupling triggered the sudden release of a massive elastic potential energy stored in the compressed west coast, resulting in the rapid restoration of the bulge-bent west and east coast deformation to the regular as an elastic rebound of the entire Tohoku. This entire elastic rebound generated the megathrust EQ and tsunami, as illustrated in Fig. 2c. Thus, the Tohoku tsunami genesis process is the east coast's pulling process.

**Precursory microgravity anomaly observations**

The continental plate-driving eastward force compressed the west coast eastward and simultaneously caused the east coast to bulge-bend, initiating the process of pulling down the subducting oceanic plate in July 2010. The plate-driving force continued to bulge-bend the east coast until the halt of pulling action in February 2011. The timeline of this pulling action by the enormous continental plate-driving force across Tohoku is consistent with a pre-seismic microgravity anomaly observed by the GRACE satellite data from July 2010 to February 2011 [9]. The upheaval growth that began in November 2010 near the Tohoku M9 epicenter area has a timeline that agrees with the uplift growth suggested by the sea-surface gravity change observation from November 2010 to February 2011 [10].

**Tsunami disaster prevention warning**

Cross-correlating Physical Wavelets (PWs) with non-differentiable {$c$} define noise-free displacement $D(c, t)$, velocity $V(c, t)$, and acceleration $A(c, t)$ at time $t$. The paths in $D(c, t) - V(c, t)$ and $D(c, t) - A(c,$ $t$) plane quantify the 2011 Tohoku EQ genesis of 15 months, having the last 3 months of disaster prevention warning [4,6]. Thus, the final phase can provide real-time information on the impending megathrust EQ and tsunami, allowing for the most effective disaster prevention warnings and hazard mitigations.

**Detecting abnormal motion**

The movements of the subducting oceanic plate, as recorded at GPS stations in the Northwest Pacific Ocean, were influenced by the lunar synodic tidal force loading. The east coast bulge-bending pulling was a new external force to the tidal loading that caused gradual changes in the amplitudes and phases of the periodic loading, which had a 30-day period. The paths of $D(E, t) – V(E, t)$ and $D(E, t) – A(E, t)$ with $PW(E, t) = V(E, t) \times A(E, t)$ can automatically detect the abnormal oceanic plate motion [4], serving as a real-time tsunami warning.

**Figures 1 and 2**

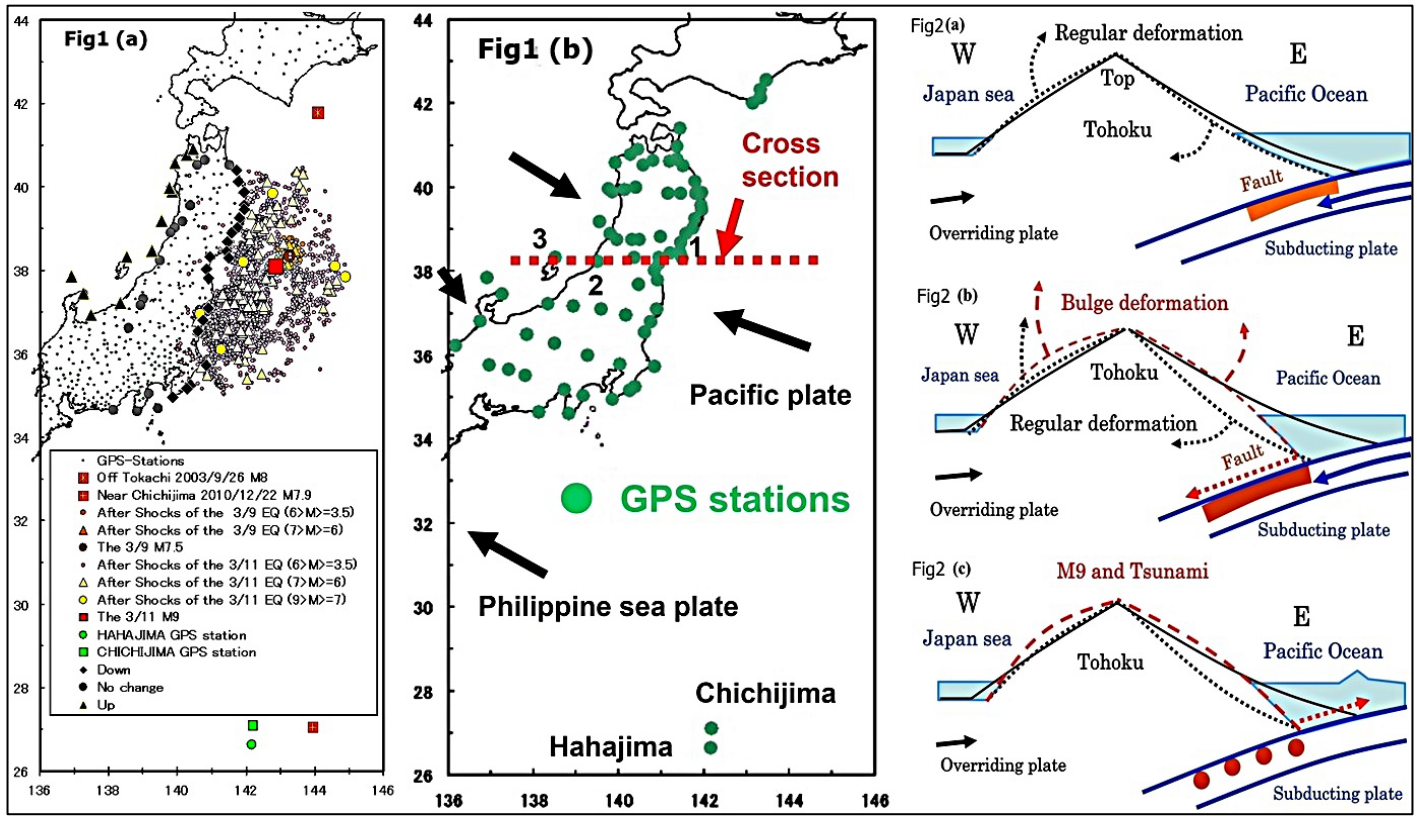

**Figure 1** displays the GPS stations in the Tohoku subduction zone, providing an extensive record of the daily position at each station. **(a)** The GPS stations recorded coseismic shifts down on the east coast, no change on TOP, and up on the west coast. **(b)** The three GPS stations, 1, 2, and 3, are referred to as the east coast (Onagawa station), the Top (Murakami station), and the west coast (Ryoutsu2 station), respectively [3, 4, 6].

**Figure 2** is a schematic cross-section at the dotted line in Fig. 1b, with stations 1, 2, and 3 referred to as Onagawa station (east coast, E), Murakami station (Top), and Ryoutsu2 station (west coast, W), respectively. **(a)** A regular slow deformation began changing the west coast deformation from non-elastic to elastic, generating a restoring force of the compressed west coast with gradual subsidence of 1 to 2.8 mm across Tohoku starting in January 2010. **(b)** The transition was from the regular to a bulge-bending deformation, which pulled the subducting Pacific Plate by coupling with the fault. As indicated by the eastward arrow, the overriding plate-driving force compressed the elastic west coast, causing the east coast to pull down the subducting plate, as indicated by a dotted arrow on the fault. The east coast's pulling action with the westward displacement began in July 2010. **(c)** The final phase began in November 2010 with the gradual upheaval growth of 1.2 mm on the east coast, generating the lifting force along the Tohoku subduction zone. The buildup of lifting force on the east coast eventually caused the shear stress to exceed the static frictional strength of the subduction interface weakened by the lifting force, causing the overriding and subducting plates to decouple. The lifting force on the entire Tohoku region helped the decoupling process, releasing a massive recoil force of the compressed west coast against the eastward-plate-driving force. The recoil rapidly restored the bulge-bent deformation of the east and west coasts elastically compressed by the overriding plate-driving eastward force. This rapid restoration led to the Tohoku M9 earthquake and tsunami on March 11, 2011 [3, 4, 6].